\documentclass[%
 reprint,
 amsmath,amssymb,
 aps,
]{revtex4-2}
\usepackage{enumitem}
\usepackage{graphicx}
\usepackage{dcolumn}
\usepackage{bm}
\usepackage{hyperref}
\usepackage{slashed}
\usepackage{subfloat}
\usepackage{multirow}
\usepackage[table]{xcolor}
\usepackage{pgfplots,subfigure}
\usepackage{orcidlink}
\usepackage{float}
\usepackage{geometry}
\usepackage[table]{xcolor} 
\usepackage{booktabs}       
\usepackage{array}          

\newcolumntype{C}{>{\centering\arraybackslash}m{4cm}} 
\definecolor{acsblue}{RGB}{17,76,139}
\definecolor{LightGreen}{RGB}{198,239,206}
\definecolor{LightRed}{RGB}{255,199,206}
\definecolor{LightGray}{RGB}{220,220,220}

\begin{document}

\fontsize{8}{9}\selectfont
\preprint{APS/123-QED}

\title{Scalar-Wave Signatures of Wormholes in Dark Matter Halos}

\author{Abdullah Guvendi\,\orcidlink{0000-0003-0564-9899}}
\email{abdullah.guvendi@erzurum.edu.tr}
\affiliation{Department of Basic Sciences, Erzurum Technical University, 25050, Erzurum, Türkiye}

\author{Omar Mustafa\,\orcidlink{0000-0001-6664-3859}}
\email{omar.mustafa@emu.edu.tr}
\affiliation{Department of Physics, Eastern Mediterranean University, 99628, G. Magusa, north Cyprus, Mersin 10 - Türkiye}

\author{Semra Gurtas Dogan\,\orcidlink{0000-0001-7345-3287}}
\email{semragurtasdogan@hakkari.edu.tr (Corresponding Author)}
\affiliation{Department of Medical Imaging Techniques, Hakkari University, 30000, Hakkari, Türkiye}

\author{Abdelghani Errehymy\,\orcidlink{0000-0002-0253-3578}}
\email{abdelghani.errehymy@gmail.com}
\affiliation{Astrophysics Research Centre, School of Mathematics, Statistics and Computer Science, University of KwaZulu-Natal, Private Bag X54001, Durban 4000, South Africa.}
\affiliation{Center for Theoretical Physics, Khazar University, 41 Mehseti Str., Baku, AZ1096, Azerbaijan}

\author{Hassan Hassanabadi\orcidlink{0000-0001-7487-6898}}
\email{hha1349@gmail.com}
\affiliation{Department of Physics, University of Hradec Kr$\acute{a}$lov$\acute{e}$, Rokitansk$\acute{e}$ho 62, 500 03 Hradec Kr$\acute{a}$lov$\acute{e}$, Czechia}
\affiliation{Department of Physics and Electronics, Khazar University, 41 Mahsati Str, 1096 Baku, Azerbaijan}

\date{\today}

\begin{abstract}
{\fontsize{8}{9}\selectfont \setlength{\parindent}{0pt}  
We identify scalar-wave signatures of massless fields propagating in static, spherically symmetric wormholes embedded within realistic dark matter halos. Starting from a general line element with arbitrary redshift and shape functions, we recast the radial Klein-Gordon equation in Schrödinger form, explicitly separating contributions from gravitational redshift, spatial curvature, and angular momentum. The dynamics reduce to a generalized Helmholtz equation with a space- and frequency-dependent effective refractive index that encodes the throat geometry, halo curvature, and centrifugal effects, asymptotically recovering free-space propagation. Applying this framework to Navarro-Frenk-White, Thomas-Fermi Bose-Einstein condensate, and Pseudo-Isothermal halo models, and considering zero, Teo-type, and cored redshift functions, we uncover evanescent regions and suppression of high-angular-momentum modes in the vicinity of the throat. High-frequency waves approach the geometric-optics regime, whereas low-frequency modes exhibit strong curvature-induced localization. In the geometric-optics limit, the effective refractive index reproduces null-geodesic trajectories, while finite-frequency effects capture evanescent zones and tunneling phenomena. This work establishes the first exact, non-perturbative framework linking wormhole geometry and realistic dark matter halos to observable scalar-wave propagation phenomena, including evanescence, mode suppression, and frequency-dependent localization.
}
\end{abstract}

\keywords{Wormholes; Dark Matter Halos; Scalar Wave Propagation; Refractive Index; Scalar Wave Optics}

\maketitle

\tableofcontents

\section{Introduction}

\setlength{\parindent}{0pt}

The concept of wormholes originates from Flamm’s 1916 analysis of the Schwarzschild geometry \cite{flamm1916}, and was later formalized by Einstein and Rosen as bridges connecting distinct spacetime regions \cite{einstein1935}. Wheeler subsequently introduced the term wormhole in the context of geons \cite{Misner:1957mt}. Although their existence remains hypothetical, wormholes arise as consistent solutions to Einstein’s equations and represent nontrivial topological connections between remote regions of the Universe. Early analyses demonstrated that generic wormhole configurations are unstable, collapsing before even photons can traverse them \cite{Klinkhamer:2023avf, Fuller:1962zza}. A major advance came with the work of Morris, Thorne, and Yurtsever \cite{morris1988, Morris:1988tu}, who showed that suitably engineered exotic matter sources can sustain traversable solutions, and that such geometries can admit closed timelike curves. Visser later introduced thin-shell constructions to minimize the required exotic matter content \cite{visser1996}. Despite these advances, violations of the null energy condition remain a generic feature of traversable wormholes. More recently, wormholes have been considered in the broader context of quantum information and the ER=EPR correspondence \cite{Marolf:2013dba}. Among traversable models, the stationary and axially symmetric Teo wormhole \cite{Teo:1998dp} represents the first explicit rotating extension of the Morris-Thorne geometry. Although it still requires exotic matter violating the null energy condition \cite{Tsukamoto:2014swa}, it provides a natural arena for studying astrophysically relevant rotating wormholes. The propagation of light and waves in such backgrounds has been extensively studied. For the Ellis wormhole, Chetouani and Clément first identified light deflection effects \cite{Chetouani:1984qdm}, and subsequent work has explored weak and strong gravitational lensing \cite{Tsukamoto:2016zdu, Tsukamoto:2012zz, Tsukamoto:2012xs, Nakajima:2012pu, Bhattacharya:2010zzb}, microlensing and retro-lensing \cite{Abe:2010ap, Tsukamoto:2017edq}, and wave-optical effects \cite{Yoo:2013cia}. Strong deflection in Janis-Newman-Winicour and Ellis wormholes has also been investigated \cite{Tsukamoto:2016qro, Dey:2008kn}, together with extensions to brane-world and scalar-tensor scenarios \cite{Nandi:2006ds, Shaikh:2017zfl}, and frame-dragging in rotating optical analogues \cite{Errehymy:2025psi}.  

\vspace{0.05cm}
\setlength{\parindent}{0pt}

Most previous studies have treated wormholes as isolated systems, yet astrophysical considerations demand embedding them in realistic galactic environments \cite{new-2025}. In particular, dark matter halos provide the dominant dynamical component of galaxies. Observations since the 1970s established that the luminous matter alone cannot account for flat galactic rotation curves at large radii \cite{Freeman:1970mx, Whitehurst:1972, Rogstad:1972, Roberts:1973}, motivating the dark matter hypothesis \cite{Rubin:1978kmz, deBlok:2002a, Walter:2008, Lelli:2016nwa}. While alternative approaches such as modified Newtonian dynamics \cite{Milgrom:1983pn, Begeman:1991, Sanders:2007, Swaters:2010} and modified gravity theories \cite{Brownstein:2005zz, Cardone:2010, Lin:2013} have been proposed, the dark matter paradigm remains the most successful framework for explaining galactic rotation curves, the growth of large-scale structure, and the cosmic microwave background. A variety of dark matter profiles have been developed to describe halo structure. The Navarro-Frenk-White (NFW) model, derived from cold dark matter $N$-body simulations, predicts cuspy inner regions \cite{Navarro:1995iw, Navarro:1996gj}, whereas the Burkert profile features a cored density suitable for dwarf galaxies \cite{burkert1995}. The pseudo-isothermal (PI) profile reproduces a wide range of rotation curves \cite{jimenez2003}, and Brownstein introduced a core-modified distribution with constant central density \cite{Brownstein:2009}. More flexible parametrizations such as the Einasto profile \cite{einasto1965} and generalized Zhao-An profiles \cite{Zhao:1995cp, An:2012pv} provide improved fits to high-precision data. These halo models are typically characterized by a density scale $\rho_s$ and a radial scale $R_s$, with additional parameters introduced to capture more complex morphologies. Probing how waves, whether electromagnetic or gravitational, propagate through such matter distributions provides a unique opportunity to uncover the optical fingerprints of dark matter halos and, even more intriguingly, to investigate how these signatures are altered in the presence of exotic spacetime structures such as traversable wormholes.

\vspace{0.05cm}
\setlength{\parindent}{0pt}

The study of wave propagation in curved spacetimes offers a powerful framework for analyzing optical phenomena in the presence of effective background potentials \cite{Rop1,Rop2}. In general relativity, curvature generated by matter distributions modifies the propagation of fields, producing effects such as wave-optical corrections to light trajectories \cite{R1}, gravitational lensing \cite{R2}, scattering and absorption by compact objects \cite{R3}, and photon ring formation near strong-gravity regions \cite{R4,R5,guvendi-npb}.
While direct detection of such phenomena is difficult due to their weak signatures, exact analytical studies provide key insights into their observable imprints. To complement astrophysical observations, laboratory analogues have been proposed to emulate curved-space wave dynamics \cite{semra-2025,R6,R7}. These rely on engineered background media to reproduce effective spacetime curvature. Examples include the observation of spontaneous Hawking-like emission in Bose-Einstein condensates \cite{R8}, simulation of Schwarzschild orbital precession with gradient-index lenses \cite{R9}, and optical analogues of gravitational lensing using microstructured waveguides \cite{R10}. Another approach employs dimensional reduction techniques, mapping curved manifolds into flat three-dimensional embeddings to facilitate controlled optical experiments \cite{R11,R12}. This strategy, pioneered by Batz and Peschel \cite{R13}, has stimulated extensive theoretical and experimental research in diverse settings \cite{R6,R14,R15}, including surface plasmon polaritons \cite{R16} and effective quantum dynamics in curved geometries \cite{R17}. Although the original Batz-Peschel analysis derived a nonlinear Schrödinger equation for waves on revolution surfaces of constant Gaussian curvature, its applicability is restricted to axisymmetric geometries and longitudinal propagation. In this work, we extend existing approaches by developing a comprehensive analytic framework for massless scalar-wave propagation in static, spherically symmetric traversable wormholes embedded in realistic dark matter halos \cite{new-2025}. We examine how halo density profiles, together with physically motivated redshift functions, shape the effective potential, the generalized refractive index, and the resulting wave dynamics. This unified treatment admits exact solutions exhibiting evanescent regions, quasi-bound states, and angular-momentum--dependent mode suppression, while consistently recovering free-space propagation at large radii. The central contribution of this work is the derivation of an exact Helmholtz formulation for scalar waves in halo-embedded wormholes, which enables a transparent refractive-index interpretation of wave propagation in realistic astrophysical environments. As a result, we establish direct and testable connections between halo morphology, wormhole geometry, and scalar-wave signatures.

\vspace{0.05cm}
\setlength{\parindent}{0pt}

The structure of this paper is as follows. Section~\ref{sec:2} derives the massless scalar wave equation in general static wormhole backgrounds and reformulates it as a generalized Helmholtz equation with a space- and frequency-dependent refractive index. Section~\ref{sec:3} introduces representative redshift functions and dark matter halo profiles, establishing analytic expressions for the effective refractive index. Section~\ref{sec:4} analyzes scalar-wave dynamics, including localization, propagation, and evanescent behavior across different halo geometries and frequency regimes. Section \ref{sec:null} establishes the direct correspondence between null geodesics and the effective refractive index, linking classical ray trajectories with scalar-wave propagation in wormhole spacetimes. Section~\ref{sec:5} summarizes our conclusions and discusses broader implications for theoretical and observational wave optics in curved spacetimes.  


\section{Exact Helmholtz Formulation of Scalar Waves in Wormholes}\label{sec:2}

\vspace{0.05cm}
\setlength{\parindent}{0pt}

We consider a static, spherically symmetric wormhole embedded in a realistic mass distribution, described by the line element \cite{new-2025}
\begin{equation}
ds^2 = -e^{2\Phi(r)} dt^2 + \frac{dr^2}{1 - \epsilon(r)/r} + r^2 \left( d\theta^2 + \sin^2\theta\, d\varphi^2 \right),\label{2.1}
\end{equation}
where $\Phi(r)$ is the redshift function, controlling gravitational time dilation, and $\epsilon(r)$ is the shape function, determining the spatial embedding of the wormhole geometry \cite{AO-2025}. The wormhole throat occurs at $r=r_0$, defined by $\epsilon(r_0)=r_0$ with the flare-out condition $\epsilon'(r_0)<1$, ensuring a non-trivial bridge between two asymptotically flat regions. These requirements generally necessitate exotic matter to satisfy the flare-out condition near the throat, although specific matter distributions or modified gravity scenarios may relax this requirement. We consider a minimally coupled, massless scalar field $\Psi(t,r,\theta,\varphi)$, whose dynamics are governed by the action
\begin{equation}
S = -\frac{1}{2} \int \sqrt{-g}\, g^{\mu\nu} \partial_\mu \Psi \, \partial_\nu \Psi \, d^4x,
\end{equation}
resulting in the curved-space massless Klein-Gordon equation \cite{konoplya}:
\begin{equation}
\Box \Psi = \frac{1}{\sqrt{-g}} \partial_\mu \left( \sqrt{-g} \, g^{\mu\nu} \, \partial_\nu \Psi \right) = 0.
\end{equation}
Exploiting spherical symmetry, we decompose the field into frequency and angular-momentum modes:
\begin{equation}
\Psi(t,r,\theta,\varphi) = e^{-i\omega t} Y_{\ell m}(\theta,\varphi) \frac{\psi(r)}{r},
\end{equation}
where $\omega$ is the mode frequency, $Y_{\ell m}(\theta,\varphi)$ are the spherical harmonics, and $\psi(r)$ is the radial function. Substituting this ansatz into the Klein-Gordon equation yields the exact radial equation:
\begin{widetext}
\begin{align}
\left(1 - \frac{\epsilon(r)}{r}\right)\psi''(r) + 
\left[\left(1 - \frac{\epsilon(r)}{r}\right)\Phi'(r) + \frac{\epsilon(r)-r\epsilon'(r)}{2r^2}\right]\psi'(r) 
+ \left[\omega^2 e^{-2\Phi(r)} - \frac{\ell(\ell+1)}{r^2} - \frac{\epsilon(r)-r\epsilon'(r)}{2 r^3} - \frac{\left(1 - \frac{\epsilon(r)}{r}\right)\Phi'(r)}{r} \right] \psi(r) = 0,
\label{eq:radial-general}
\end{align}
\end{widetext}
where a prime denotes a derivative with respect to $r$. It is advantageous to transform equation \eqref{eq:radial-general} into the canonical one-dimensional Schrödinger form
{(i.e., Helmholtz equation form)} by performing the field redefinition
\begin{equation}
\psi(r) = f(r)\,\chi(r), \quad f(r) = e^{-\Phi(r)/2} \left(1 - \frac{\epsilon(r)}{r}\right)^{-1/4},
\end{equation}
which removes the first-order derivative term and normalizes the second-derivative term coefficient, yielding a canonical one-dimensional Schrödinger-like equation. The radial dynamics then assume the Schrödinger-type Helmholtz form
\begin{equation}
\chi^{''} + \left[\omega^2 e^{-2\Phi(r)} + \frac{\epsilon''(r)}{4r} + \frac{3\,(r\epsilon'(r)-\epsilon(r))^2}{16 r^3 (r-\epsilon(r))} - \frac{\ell(\ell+1)}{r^2}\right]\chi = 0,
\label{eq:schrodinger-type}
\end{equation}
where the effective potential distinctly separates the influences of gravitational redshift, spatial curvature, and angular momentum, establishing a rigorous framework for both analytical and numerical investigations of scalar-wave propagation in spherically symmetric wormhole geometries. The centrifugal barrier $-\ell(\ell+1)/r^2$ appears explicitly, whereas curvature effects arising from the halo are encoded in the shape function and its derivatives. This formulation can be expressed in the generalized Helmholtz form \cite{semra-2025,PLB-2,EPJC-1,Ahmed-2025}
\begin{equation}
\chi^{''}(r) + \omega^2\, n_{\mathrm{eff}}^2(r,\omega) \, \chi(r) = 0,\label{2.2}
\end{equation}
with a frequency- and position-dependent effective refractive index \cite{semra-2025}
\begin{equation}
n_{\mathrm{eff}}^2(r,\omega) = e^{-2\Phi(r)} + \frac{1}{\omega^2} \left[ \frac{\epsilon''(r)}{4 r} + \frac{3 \left(r\epsilon'(r) - \epsilon(r)\right)^2}{16 r^3 (r-\epsilon(r))} - \frac{\ell(\ell+1)}{r^2} \right].\label{Ref-index}
\end{equation}
The first term, $e^{-2\Phi(r)}$, encodes gravitational redshift, whereas the remaining contributions capture curvature-induced corrections and centrifugal suppression of higher-angular-momentum modes. In the asymptotic limit $r \to \infty$, one has $\epsilon(r)/r \to 0$ and $\Phi(r) \to 0$, so that $n_{\mathrm{eff}} \to 1$, recovering free-space propagation. In the high-frequency regime $\omega \to \infty$, derivative contributions become negligible, yielding $n_{\mathrm{eff}}^2 \simeq e^{-2\Phi(r)}$, consistent with the geometric-optics limit. Conversely, low-frequency modes are strongly influenced by curvature and may exhibit evanescent behavior or localization near the wormhole throat. This exact Helmholtz formulation provides a fully analytic framework for scalar-wave propagation in wormholes immersed in dark matter halos, while facilitating direct analogies with optical systems and efficient numerical implementation.

\begin{figure*}[!ht]
\centering
\includegraphics[scale=0.75]{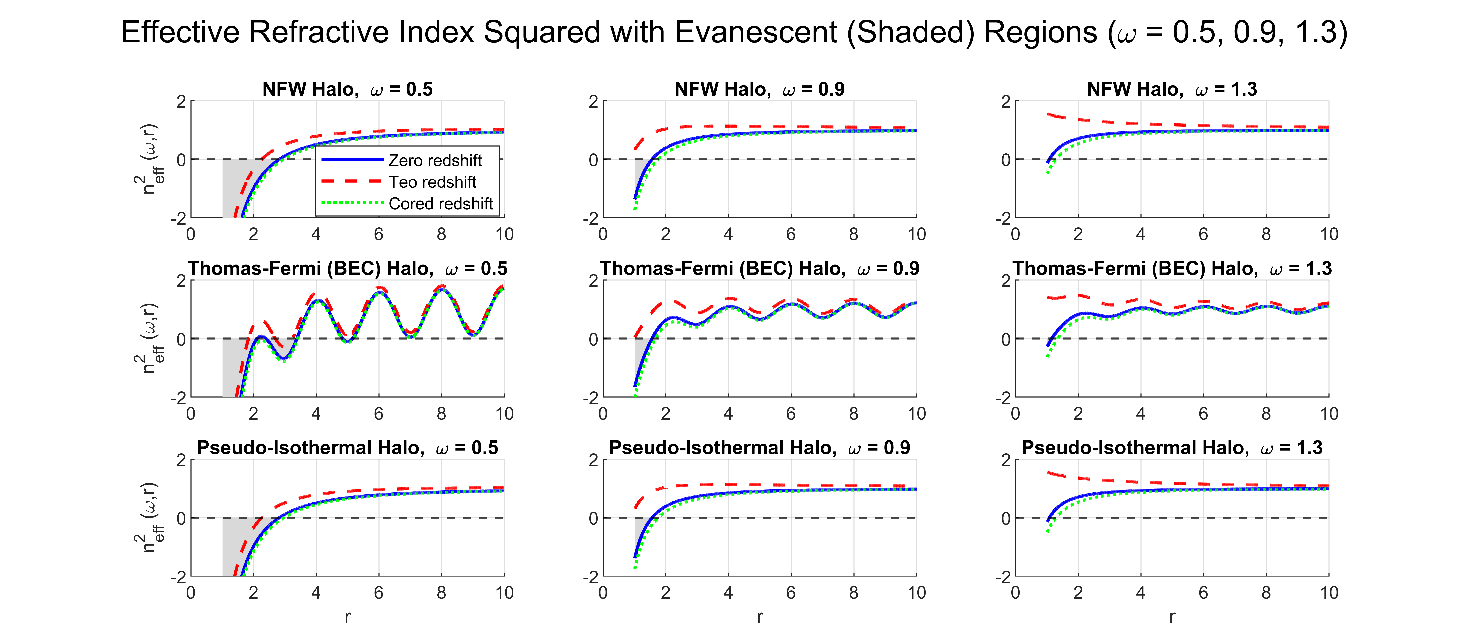}
\caption{\fontsize{8}{9}\selectfont Effective refractive index squared $n_\mathrm{eff}^2(\omega,r)$ for various wormhole and halo models as a function of the radial coordinate $r$. The rows correspond to different halo profiles: (i) NFW Halo, (ii) TF (BEC) Halo, and (iii) PI Halo. The columns represent different frequencies $\omega = 0.5, 0.9, 1.3$. Redshift functions considered are: Zero redshift $\Phi(r) = 0$, Teo redshift $\Phi(r) = -a/r$ with $a = 0.5$, and Cored redshift $\Phi(r) = \ln\left[ 1 + A / (1 + (r/r_0)^{n_\mathrm{Core}} ) \right]$ with $A = 0.5$, $r_0 = 1.0$ and $n_\mathrm{Core} = 2$. Shape functions for the halos are parametrized using $\rho_s = 0.01$, $R_s = 1.0$, and $\epsilon_0 = 0.1$. Shaded gray regions indicate evanescent zones where $n_\mathrm{eff}^2 < 0$, corresponding to classically forbidden regions with exponential wave decay. The black dashed line marks $n_\mathrm{eff}^2 = 0$.}
\label{fig:refractive-index}
\end{figure*}

\section{Representative Redshift Functions and Dark Matter Halo Profiles}\label{sec:3}

\vspace{0.05cm}
\setlength{\parindent}{0pt}

The redshift function $\Phi(r)$ governs the local modulation of the scalar-field frequency and directly affects the effective refractive index $n_{\rm eff}(r,\omega)$. A Teo-type redshift, $\Phi(r) = -a/r$, enhances the local frequency near the throat, thereby reducing the spatial extent of evanescent regions and increasing the density of quasi-bound states. Cored halo-inspired redshift profiles,
\begin{equation}
\Phi(r) = \ln \Bigg[ 1 + \frac{A}{1 + (r/r_0)^n} \Bigg], \quad A>-1, \ n\ge 1,
\end{equation}
introduce a gradual radial modulation at intermediate distances, reflecting physically realistic finite-core halo structures. The constraint \(A>-1\) guarantees that the redshift function remains finite and real for all \(r\), ensuring a strictly positive lapse function \(e^{2\Phi(r)}\), the preservation of a Lorentzian metric signature, the absence of horizon formation, and a well-defined effective refractive index governing scalar-wave propagation. The requirement \(n \ge 1\) ensures a smooth radial decay with finite derivatives, thereby avoiding singular curvature contributions near the throat and preserving a well-defined Helmholtz (Schrödinger-type) formulation for scalar-wave dynamics. Zero-redshift profiles, $\Phi(r) = 0$, isolate the purely geometric effects of the shape function \cite{AO-2025}. The shape function $\epsilon(r)$ encodes the radial mass distribution of the dark matter halo and determines curvature-induced contributions to $n_{\rm eff}^2$ through its derivatives. NFW halos feature a central cusp that produces steep radial gradients, enhancing curvature near the throat and generating sharply localized quasi-bound states at low frequencies. TF (BEC) halos exhibit oscillatory behavior in $\epsilon(r)$, producing multiple quasi-bound regions with smoother spatial variations, while PI halos are characterized by a monotonic, smooth shape function that yields a nearly uniform refractive index, allowing nearly free propagation over a broad frequency range. In all cases, the asymptotic condition $\epsilon(r)/r \to 0$ ensures $n_{\rm eff}^2 \to e^{-2\Phi(r)}$, recovering free-space propagation far from the throat. Inserting these representative halo shape functions and their derivatives \cite{new-2025} into the exact expression \eqref{Ref-index} for $n_{\rm eff}^2(r,\omega)$ provides a fully analytic framework for scalar-wave propagation in wormholes embedded in realistic dark matter halos. This formulation systematically combines the contributions of throat geometry, halo-induced curvature, and gravitational redshift without relying on perturbative approximations, enabling precise analytical and numerical investigations. Specifically, for the NFW profile, $\epsilon'(r) \to 0$ and $\epsilon''(r)$ remains finite near the origin, while both derivatives vanish at large $r$, capturing the cuspy central density and asymptotic falloff. The TF (BEC) halo exhibits $\epsilon'(r) \to 0$ and $\epsilon''(r) \to 0$ at the origin, with oscillations around $r \sim R_s$, characteristic of a finite-size condensate. In the PI profile, both derivatives vanish at $r = 0$, while $\epsilon'(r)$ approaches a constant and $\epsilon''(r) \to 0$ at large radii, reflecting a flattened outer halo. These behaviors confirm the absence of central singularities and provide clear insight into the asymptotic structure of each halo model.

\begin{table*}[!htbp]
\centering
\fontsize{8}{9}\selectfont
\renewcommand{\arraystretch}{1.3}
\caption{ \fontsize{8}{9}\selectfont Numerically determined radial turning points $r$ at which the effective refractive index satisfies $n_{\rm eff}^2(r,\omega)=0$ for different dark matter halo profiles (NFW, TF, and PI) and redshift prescriptions
(Zero, Teo, and Cored). For each frequency $\omega$, the corresponding number of evanescent zones, defined as
radial intervals where $n_{\rm eff}^2<0$ and wave propagation is forbidden, is reported. The background color (orange)
intensity represents the number of evanescent zones, with darker shading indicating a larger number of
non propagating regions.}
\vspace{0.5cm}
\begin{tabular}{@{}llccccccc@{}}
\toprule
\rowcolor{gray!20} 
\textbf{Halo} & \textbf{Redshift} & $\omega=0.5$ & Zones & $\omega=0.9$ & Zones & $\omega=1.3$ & Zones \\
\midrule
NFW & Zero  & 2.82983  & \cellcolor{orange!20}1 & 1.56781 & \cellcolor{orange!20}1 & 1.08004 & \cellcolor{orange!20}1 \\
NFW & Teo   & 2.26864  & \cellcolor{orange!20}1 & --      & \cellcolor{white}0 & --      & \cellcolor{white}0 \\
NFW & Cored & 2.97387  & \cellcolor{orange!20}1 & 1.76053 & \cellcolor{orange!20}1 & 1.28703 & \cellcolor{orange!20}1 \\
TF  & Zero  & 2.11541, 2.39500, 3.38402, 4.84014, 5.16014 & \cellcolor{orange!50}3 & 1.56028 & \cellcolor{orange!20}1 & 1.15355 & \cellcolor{orange!20}1 \\
TF  & Teo   & 1.84174, 2.69184, 3.23858 & \cellcolor{orange!30}2 & --      & \cellcolor{white}0 & --      & \cellcolor{white}0 \\
TF  & Cored & 3.41105, 4.81183, 5.18474 & \cellcolor{orange!30}2 & 1.67501 & \cellcolor{orange!20}1 & 1.34643 & \cellcolor{orange!20}1 \\
PI  & Zero  & 2.81925 & \cellcolor{orange!20}1 & 1.56108 & \cellcolor{orange!20}1 & 1.07931 & \cellcolor{orange!20}1 \\
PI  & Teo   & 2.25618 & \cellcolor{orange!20}1 & --      & \cellcolor{white}0 & --      & \cellcolor{white}0 \\
PI  & Cored & 2.96401 & \cellcolor{orange!20}1 & 1.75373 & \cellcolor{orange!20}1 & 1.28369 & \cellcolor{orange!20}1 \\
\bottomrule
\end{tabular}
\label{tab:tp}
\end{table*}

\section{Scalar-Wave Dynamics and Effective Refractive Index in Wormholes} \label{sec:4}

\vspace{0.05cm}
\setlength{\parindent}{0pt}

Figure~\ref{fig:refractive-index} shows the radial and frequency dependence of the effective refractive index squared $n_\mathrm{eff}^2(\omega,r)$ for distinct halo density profiles and representative redshift functions. In all cases, particularly for low-frequency waves, $n_\mathrm{eff}^2$ decreases near the wormhole throat at $r_0 = 1.0$ and can become negative, generating evanescent regions where wave propagation is forbidden and fields decay exponentially. The zero-redshift configuration serves as a baseline, isolating the effects of the halo-induced geometry. Teo-type and cored redshifts modify the refractive index profile in contrasting ways: the Teo potential induces a pronounced drop near the throat due to its $1/r$ scaling, while the cored redshift generates a smoother radial variation, reducing the spatial extent of evanescent zones. Increasing the wave frequency $\omega$ systematically shrinks the evanescent regions, in agreement with the $\omega^{-2}$ scaling in $n_\mathrm{eff}^2$. The effective refractive index $n_\mathrm{eff}(r,\omega)$ serves as a direct optical analogue for scalar-wave propagation in the wormhole-halo system. Near the throat, the flare-out condition enforces a minimal areal radius, causing curvature terms proportional to $\epsilon(r) - r\epsilon'(r)$ and $\epsilon''(r)$ to dominate. High-frequency modes undergo rapid radial oscillations, characteristic of the geometric optics regime, whereas low-frequency waves are strongly affected by curvature, forming evanescent regions and quasi-bound states confined near the throat. The centrifugal barrier $\ell(\ell+1)/r^2$ (i.e., central repulsive force field core) explicitly suppresses higher angular-momentum modes, effectively filtering the mode spectrum. At intermediate radial distances, the behavior of $n_\mathrm{eff}(r,\omega)$ is highly sensitive to the halo profile. NFW halos, with steep central cusps, amplify the refractive index sharply, producing highly localized quasi-bound states, while high-frequency waves propagate with minimal curvature effects. TF (BEC) halos introduce oscillatory modulations, creating extended quasi-bound regions interspersed with multiple evanescent zones. PI halos, with their smooth and cored distribution, maintain an almost uniform refractive index, allowing nearly free propagation across a broad frequency range. In the asymptotic limit $r \to \infty$, all halo profiles satisfy $n_\mathrm{eff}^2 \to e^{-2\Phi(r)}$, recovering free-space propagation in the weak-field regime. The redshift function modulates the local effective frequency, shifting the radial turning points where $n_\mathrm{eff}^2$ crosses zero. Due to the transcendental structure of the Eq.~\eqref{Ref-index}, arising from the combined dependence on $\epsilon(r)$, its derivatives, and the redshift function $\Phi(r)$, analytical solutions for the turning points cannot be obtained in general. Closed expressions are accessible only in limiting regimes such as the near throat region or the asymptotic large radius limit, where the structure of $n_{\rm eff}^2(r,\omega)$ simplifies. At intermediate radii, where curvature and redshift effects compete, the equation remains non algebraic. For this reason, all turning points reported in Table~\ref{tab:tp} are determined numerically, ensuring that the full structure of Eq.~\eqref{Ref-index} is retained without invoking asymptotic or perturbative approximations. Our results demonstrate that Teo-type redshift functions compress evanescent regions and significantly enhance the density of quasi-bound states, whereas cored redshifts yield smoother spectral transitions. In the high-frequency regime, derivative contributions to $n_{\mathrm{eff}}^{2}$ become negligible, and the wave dynamics asymptotically approach the predictions of geometric optics. By contrast, low-frequency modes are strongly influenced by spacetime curvature, exhibiting pronounced localization, evanescent attenuation, and the possible emergence of quasi-normal-mode echoes. 

\vspace{0.05cm}
\setlength{\parindent}{0pt}

As summarized in Table~\ref{tab:tp}, the numerical turning points at which the effective refractive index
satisfies $n_{\rm eff}^2(r,\omega)=0$ exhibit a clear dependence on both the halo density profile and the
redshift prescription. For all halo models considered, increasing the frequency $\omega$ shifts the turning
points toward smaller radial distances and systematically reduces the number of evanescent zones. This
trend indicates that higher frequency modes are progressively less affected by the effective potential
induced by the gravitational redshift and halo geometry, thereby allowing wave propagation to penetrate
closer to the central region. Table~\ref{tab:tp} further shows that the NFW and PI halo profiles predominantly support a single turning point over the explored frequency range, corresponding to a single evanescent barrier separating
propagating and non propagating regions. In contrast, the TF halo profile exhibits multiple turning points
at low frequency, leading to the formation of two or more distinct evanescent zones. This behavior is a
direct consequence of the extended and oscillatory structure of the TF density profile, which generates
multiple changes in the effective refractive index. The influence of the redshift prescription is also evident in Table~\ref{tab:tp}. The Teo redshift suppresses the formation of turning points at higher frequencies and eliminates evanescent zones for $\omega \ge 0.9$, indicating reduced gravitational confinement and uninterrupted propagation. Conversely, the Cored redshift preserves turning points across all examined frequencies, reflecting stronger central confinement and a
more persistent refractive barrier. Overall, the results presented in Table~\ref{tab:tp} demonstrate that
the location and multiplicity of evanescent zones provide a sensitive diagnostic of both the underlying
halo structure and the spacetime redshift, with important implications for wave trapping and attenuation in
curved astrophysical environments.

\section{Correspondence Between Null Geodesics and the Effective Refractive Index}
\label{sec:null}

The propagation of massless particles in a curved spacetime is governed by null geodesics, which may be derived most transparently within the Hamilton-Jacobi formalism. For a massless particle, Hamilton’s principal function $S(x^\mu)$ satisfies the null Hamilton-Jacobi equation 
\begin{equation}
g^{\mu\nu}\,\partial_\mu S\,\partial_\nu S = 0 ,
\label{HJ}
\end{equation}
where $g^{\mu\nu}$ denotes the inverse spacetime metric. We consider a static, spherically symmetric wormhole spacetime described by the line element \eqref{2.1}, where $\Phi(r)$ is the redshift function and $\epsilon(r)$ is the shape function. Owing to spherical symmetry, the motion of null particles may be confined, without loss of generality, to the equatorial plane $\theta=\pi/2$. The metric admits two Killing vectors, $\partial_t$ and $\partial_\phi$, associated respectively with time-translation invariance and axial rotational symmetry. As a consequence, the energy $\mathcal{E}$ and angular momentum $\mathcal{L}$ of the null particle are conserved along the trajectory. These symmetries allow the Hamilton-Jacobi function to be separated in the form
\begin{equation}
S=-\mathcal{E}t+\mathcal{L}\phi+S_r(r),
\label{action}
\end{equation}
where $S_r(r)$ depends only on the radial coordinate. The canonical momenta,
\begin{equation}
p_\mu=\frac{\partial S}{\partial x^\mu},
\end{equation}
are therefore
\begin{equation}
p_t=-\mathcal{E},\qquad
p_\phi=\mathcal{L},\qquad
p_r=\frac{dS_r}{dr}.
\end{equation}
The nonvanishing components of the inverse metric corresponding to Eq.~\eqref{2.1} are
\begin{equation}
g^{tt}=-e^{-2\Phi(r)},\qquad
g^{rr}=1-\frac{\epsilon(r)}{r},\qquad
g^{\phi\phi}=\frac{1}{r^2}.
\end{equation}
Substituting the separated action \eqref{action} into the Hamilton-Jacobi equation \eqref{HJ} yields
\begin{equation}
-e^{-2\Phi(r)}\mathcal{E}^2
+\left(1-\frac{\epsilon(r)}{r}\right)p_r^2
+\frac{\mathcal{L}^2}{r^2}=0 .
\end{equation}
Solving for the radial momentum squared gives the following expression
\begin{equation}
p_r^2
=\frac{1}{1-\epsilon(r)/r}
\left(
\mathcal{E}^2e^{-2\Phi(r)}-\frac{\mathcal{L}^2}{r^2}
\right),
\label{pr2}
\end{equation}
which completely characterizes the radial dynamics of null geodesics in this geometry. Physical motion requires $p_r^2\ge0$, and radial turning points are therefore determined by
\begin{equation}
\mathcal{E}^2e^{-2\Phi(r)}=\frac{\mathcal{L}^2}{r^2}.
\label{tp}
\end{equation}
Introducing the impact parameter $b\equiv\mathcal{L}/\mathcal{E}$, this condition becomes
\begin{equation}
b^2=r^2e^{-2\Phi(r)}.
\label{impact}
\end{equation}
Depending on the value of $b$ and the functional form of $\Phi(r)$, this equation may admit zero, one, or multiple real solutions, corresponding respectively to complete transmission through the wormhole, critical trajectories, or reflection back to infinity. Owing to the nontrivial spatial topology of wormhole geometries, multiple turning points may occur even when the redshift function is monotonic. Unstable circular null orbits arise when the turning point becomes degenerate, which occurs when the impact parameter attains an extremum as a function of $r$. Differentiating Eq.~\eqref{impact} with respect to $r$ yields the photon-sphere condition
\begin{equation}
\frac{d}{dr}\!\left(r^2e^{-2\Phi(r)}\right)=0 .
\label{CNO}
\end{equation}
Solutions of this equation determine the radius $r_{\rm ph}$ of circular null orbits. Depending on the behavior of $\Phi(r)$ and $\epsilon(r)$, wormhole spacetimes may also admit circular null orbits located directly at the throat. The corresponding critical impact parameter is
\begin{equation}
b_c^2=\left.r^2e^{-2\Phi(r)}\right|_{r=r_{\rm ph}} .
\end{equation}
The spacetime trajectories follow from Hamilton’s equations,
\begin{equation}
\frac{dx^\mu}{d\tau}=g^{\mu\nu}p_\nu ,
\end{equation}
where $\tau$ is an affine parameter along the null geodesic. The radial equation of motion becomes
\begin{equation}
\frac{dr}{d\tau}
=\left(1-\frac{\epsilon(r)}{r}\right)p_r,
\end{equation}
which, upon substituting Eq.~\eqref{pr2}, yields
\begin{equation}
\left(\frac{dr}{d\tau}\right)^2
=\left(1-\frac{\epsilon(r)}{r}\right)
\left(
\mathcal{E}^2e^{-2\Phi(r)}-\frac{\mathcal{L}^2}{r^2}
\right).
\label{rdot}
\end{equation}
The azimuthal equation of motion is
\begin{equation}
\frac{d\phi}{d\tau}
=g^{\phi\phi}p_\phi
=\frac{\mathcal{L}}{r^2}.
\end{equation}
Combining this relation with Eq.~\eqref{rdot} gives
\begin{equation}
\frac{d\phi}{dr}
=
\frac{\mathcal{L}}{
\sqrt{\left(1-\epsilon(r)/r\right)
\left(\mathcal{E}^2e^{-2\Phi(r)}-\frac{\mathcal{L}^2}{r^2}\right)}
}.
\end{equation}
The total deflection angle for a null trajectory with impact parameter $b$ is therefore
\begin{equation}
\Delta\phi(b)
=
2\int_{r_{\rm tp}}^{\infty}
b\,dr
\left[
\left(1-\frac{\epsilon(r)}{r}\right)
\left(e^{-2\Phi(r)}-\frac{b^2}{r^2}\right)
\right]^{-1/2}
-\pi .
\end{equation}
It is often convenient to rewrite the radial equation of motion in an effective-potential form,
\begin{equation}
\left(\frac{dr}{d\tau}\right)^2 + V_{\rm eff}^2(r)=\mathcal{E}^2 ,
\end{equation}
from which the effective potential governing null geodesics is identified as
\begin{equation}
V_{\rm eff}(r)=\frac{\mathcal{L}}{r}\,e^{\Phi(r)} .
\end{equation}
Circular null orbits are characterized by the simultaneous conditions $\mathcal{E}=V_{\rm eff}$ and $dV_{\rm eff}/dr=0$, which reproduce exactly the photon-sphere condition \eqref{CNO}. The correspondence with scalar-wave propagation becomes manifest in the high-frequency, or geometric-optics, limit. After separation of variables, the scalar field equation reduces to a radial Helmholtz-type equation,
\begin{equation}
\chi^{''}+k_r^2\,(r,\omega)\chi=0,
\end{equation}
where $r$ is the radial coordinate and the local radial wave number is defined by \(k_r^2(r,\omega)=\omega^2 n_{\rm eff}^2(r,\omega)\). Wave turning points are determined by $k_r^2=0$, or equivalently $n_{\rm eff}^2=0$. In the limit $\omega\to\infty$ with $\ell/\omega$ held fixed, all curvature- and derivative-dependent terms in $n_{\rm eff}^2$ are suppressed by powers of $1/\omega$, and the turning-point condition reduces to
\begin{equation}
e^{-2\Phi(r)}=\frac{\ell(\ell+1)}{\omega^2 r^2}.
\end{equation}
Identifying
\begin{equation}
b^2=\lim_{\omega\to\infty}\frac{\ell(\ell+1)}{\omega^2},
\end{equation}
this condition coincides precisely with the null-geodesic turning-point relation \eqref{impact} (with $\omega\to\mathcal{E}$). Moreover, extrema of $n_{\rm eff}^2$ reproduce the photon-sphere condition \eqref{CNO}, establishing the exact correspondence between scalar-wave and null-geodesic dynamics in the geometric-optics regime. More generally, one finds
\begin{equation}
k_r^2(r,\omega)
\;\longrightarrow\;
\left(1-\frac{\epsilon(r)}{r}\right)p_r^2 ,
\end{equation}
demonstrating that the effective refractive index reproduces the full null radial Hamiltonian in the high-frequency limit, while finite-frequency corrections encode genuinely wave-optical effects such as evanescent regions, quasi-bound states, and tunneling phenomena absent from the purely classical description.

\subsection{Photon Spheres and Shadow Formation for Representative Redshift Profiles}\label{subsec:shadow}

The existence of photon spheres in static, spherically symmetric wormhole spacetimes is governed by the extremization of the function
\begin{equation}
\mathcal{F}(r) \equiv r^{2} e^{-2\Phi(r)},
\end{equation}
which controls both the degeneracy of radial turning points for null geodesics and the trapping of high-frequency waves in the geometric-optics limit. A photon sphere exists if and only if $\mathcal{F}(r)$ admits a stationary point at a finite radius $r_{\mathrm{ph}}$ lying in the physical domain $r \ge r_{0}$, where $r_{0}$ denotes the throat radius. This condition depends exclusively on the redshift function $\Phi(r)$ and is entirely independent of the shape function. For the zero-redshift configuration $\Phi(r)=0$, the trapping function reduces identically to \(\mathcal{F}(r)=r^{2}\). Its first derivative is \(\frac{d\mathcal{F}}{dr}=2r\), which is strictly positive for all $r>0$, while the second derivative, \(\frac{d^{2}\mathcal{F}}{dr^{2}}=2\), is constant and positive. Hence, $\mathcal{F}(r)$ is strictly monotonic and possesses no extremum at any finite radius. The absence of a stationary point excludes the existence of circular null orbits. All null trajectories therefore either traverse the wormhole throat or escape to infinity, implying that no critical impact parameter exists and the spacetime casts no shadow. This conclusion holds irrespective of the choice of shape function and reflects the complete optical transparency of zero-redshift wormholes. For the Teo redshift profile $\Phi(r)=-a/r$ with $a>0$, the trapping function becomes
\begin{equation}
\mathcal{F}(r)=r^{2}e^{2a/r}.
\end{equation}
Differentiation yields
\begin{equation}
\frac{d\mathcal{F}}{dr}
=
\frac{d}{dr}\!\left(r^{2}e^{2a/r}\right)
=
2re^{2a/r}
-
2ae^{2a/r}
=
2e^{2a/r}(r-a).
\end{equation}
The extremization condition $d\mathcal{F}/dr=0$ admits the unique solution
\begin{equation}
r_{\mathrm{ph}}=a.
\end{equation}
The second derivative is
\begin{equation}
\frac{d^{2}\mathcal{F}}{dr^{2}}
=
2e^{2a/r}\left(1-\frac{2a}{r}\right),
\end{equation}
which, when evaluated at $r=a$, gives
\begin{equation}
\left.\frac{d^{2}\mathcal{F}}{dr^{2}}\right|_{r=a}
=
-\frac{2e^{2}}{a}<0.
\end{equation}
Thus, the extremum corresponds to an unstable circular null orbit (see Table \ref{tab:photon_spheres}). The associated critical impact parameter is
\begin{equation}
b_{c}^{2}
=
\mathcal{F}(r_{\mathrm{ph}})
=
a^{2}e^{2}, \qquad \text{where} \qquad e=\exp(1).
\end{equation}
Null rays with $b<b_{c}$ cross the photon sphere and traverse the wormhole, whereas those with $b>b_{c}$ are scattered back to infinity. The boundary $b=b_{c}$ defines a sharp shadow edge for asymptotic observers. The exponential dependence of $b_{c}$ on $a$ implies a rapid growth of the shadow radius with increasing redshift strength, providing a distinctive signature of Teo-type wormholes. We now turn to the cored redshift profile
\begin{equation}
\Phi(r)=\ln\!\left[1+\frac{A}{1+(r/r_{0})^{n}}\right],
\qquad A>-1,\quad n\geq 1,
\end{equation}
where the condition $A>-1$ ensures regularity of the metric and the positivity of $e^{2\Phi(r)}$ throughout the spacetime. The trapping function takes the form
\begin{equation}
\mathcal{F}(r)
=
\frac{r^{2}}{\left(1+\dfrac{A}{1+(r/r_{0})^{n}}\right)^{2}}.
\end{equation}
Differentiating explicitly yields
\begin{widetext}
\begin{equation}
\frac{d\mathcal{F}}{dr}
=
\frac{2r}{\left(1+\dfrac{A}{1+(r/r_{0})^{n}}\right)^{2}}
-
\frac{2r^{2}A n (r/r_{0})^{n-1}}
{r_{0}\left(1+(r/r_{0})^{n}\right)^{2}
\left(1+\dfrac{A}{1+(r/r_{0})^{n}}\right)^{3}}.
\end{equation}
\end{widetext}
For $A>0$, the first term is strictly positive and dominates the second, which is suppressed by an additional factor of $(1+(r/r_{0})^{n})^{-1}$. Consequently,
\begin{equation}
\frac{d\mathcal{F}}{dr}>0
\qquad \forall\, r>0,
\end{equation}
and $\mathcal{F}(r)$ is strictly monotonic. No photon sphere exists in this regime. For negative core strengths $-1<A<0$, the extremization condition must be solved exactly. Introducing the dimensionless variable
\begin{equation}
\rho=\left(\frac{r}{r_{0}}\right)^{n},
\end{equation}
the condition $d\mathcal{F}/dr=0$ reduces to the quadratic equation
\begin{equation}
\rho^{2}+[2+(n+1)A]\rho+(1+A)=0.
\end{equation}
Real solutions require a nonnegative discriminant,
\begin{equation}
\Delta
=
[2+(n+1)A]^{2}-4(1+A)
=
A\,[A(n+1)^{2}+4n].
\end{equation}
For $-1<A<0$, this implies
\begin{equation}
A\le -\frac{4n}{(n+1)^{2}}.
\end{equation}
Combining this with the regularity condition yields the admissible interval
\begin{equation}
-1<A\le -\frac{4n}{(n+1)^{2}}.
\end{equation}
A physical photon sphere must furthermore lie outside the throat, $r_{\mathrm{ph}}>r_{0}$, which requires $\rho\ge1$. Setting $\rho=1$ in the quadratic equation gives
\begin{equation}
1+[2+(n+1)A]+(1+A)=0,
\end{equation}
from which one obtains
\begin{equation}
A=-\frac{4}{n+2}.
\end{equation}
Therefore, a photon sphere outside the throat would require
\begin{equation}
-1<A\le \min\!\left(-\frac{4n}{(n+1)^{2}},-\frac{4}{n+2}\right).
\end{equation}
For $n=1$ and $n=2$, no value of $A>-1$ satisfies this condition. For all $n\ge3$, the extremum is located at $\rho\le1$, corresponding to a marginal or interior orbit at the throat. Consequently, no physically acceptable cored redshift wormhole admits a photon sphere outside the throat for any allowed value $A>-1$. From the wave-optical perspective, identical conclusions follow. In the geometric-optics limit, the condition $k_{r}^{2}=0$ coincides exactly with the null-geodesic turning-point condition, and extrema of the effective refractive index reproduce the photon-sphere criterion. Finite-frequency effects may introduce evanescent regions and tunneling but cannot generate photon spheres in the absence of a classical extremum of $\mathcal{F}(r)$. These results demonstrate unequivocally that photon-sphere formation and shadow production in wormhole spacetimes are dictated by the detailed radial structure of the redshift function rather than topology alone. Among the profiles examined here, only the Teo redshift supports unstable null trapping and an "observable" shadow, whereas zero-redshift and cored redshift configurations, including those with negative but regular cores, remain generically transparent and shadowless.

\begin{table}[h!]
\centering
\caption{\fontsize{8}{9}\selectfont Photon sphere formation for representative redshift profiles in static, spherically symmetric wormholes. Zero-redshift (\(\Phi=0\)) and cored profiles (\(A>-1\)) are strictly monotonic and do not admit photon spheres, making the spacetime transparent and shadowless. The Teo profile (\(\Phi=-a/r\)) admits an unstable photon sphere at \(r_\mathrm{ph}=a\) with critical impact parameter \(b_c = a e\), producing a sharp shadow.}
\label{tab:photon_spheres}
\vspace{0.3cm}
{\fontsize{8}{9}\selectfont
\begin{tabular}{@{}lll@{}}
\toprule
\rowcolor{LightGray}
\textbf{Redshift Profile} & \textbf{Derivative \(d\mathcal{F}/dr\)} & \textbf{Photon Sphere} \\
\midrule
Zero-redshift, \(\Phi = 0\) & \(2r > 0\) & \cellcolor{LightRed}No \\
Teo, \(\Phi = -a/r\) & \(2 e^{2a/r} (r-a)\) & \cellcolor{LightGreen}Yes, \(r_\mathrm{ph}=a\) \\
Cored, \(A>0\) & Positive for all \(r>0\) & \cellcolor{LightRed}No \\
Cored, \(-1<A<0\) & Quadratic in \(\rho=(r/r_0)^n\) & \cellcolor{LightRed}No \\
\bottomrule
\end{tabular}}
\end{table}


\section{Summary and Discussion}\label{sec:5}

\vspace{0.05cm}
\setlength{\parindent}{0pt}

In this manuscript, we developed a fully analytic, non-perturbative framework to study minimally coupled, massless scalar fields propagating in static, spherically symmetric wormholes embedded in realistic dark matter halos. Beginning with the general metric (\ref{2.1}), we derived the exact radial Klein-Gordon equation and cast it into a generalized Helmholtz-like form (in units where $c=1$) as reported in \eqref{2.2} and \eqref{Ref-index}. This formulation separates the contributions of gravitational redshift, halo-induced curvature, and centrifugal angular-momentum effects. The effective refractive index $n_{\rm eff}(r,\omega)$ provides a direct optical analogy, governing propagation, mode localization, and evanescent decay in the vicinity of the throat and across the halo.

\vspace{0.05cm}
\setlength{\parindent}{0pt}

Our analysis of representative halo models, including NFW, TF (BEC), and PI profiles, in combination with zero, Teo-type, and cored redshift functions, reveals that $n_{\rm eff}(r,\omega)$ is highly sensitive to both the halo morphology and the redshift structure. Near the throat, the flare-out condition enforces a minimal areal radius, amplifying curvature contributions and giving rise to evanescent regions and quasi-bound states at low frequencies. High-frequency modes propagate with minimal scattering and asymptotically reach the geometric optics limit, whereas the centrifugal barrier selectively suppresses higher angular-momentum modes, refining the mode spectrum.

\vspace{0.05cm}
\setlength{\parindent}{0pt}

At intermediate and asymptotic distances, the halo density profile dictates the spatial modulation of $n_{\rm eff}$. Cuspy NFW halos generate sharply localized quasi-bound regions, TF halos produce multiple oscillatory zones, and PI halos maintain nearly uniform propagation. Redshift functions modulate the local effective frequency, compressing or smoothing evanescent regions according to their functional form. Asymptotically, $\epsilon(r)/r \to 0$ and $n_{\rm eff}^2 \to e^{-2\Phi(r)}$, recovering free-wave propagation in the weak-field limit.

\vspace{0.05cm}
\setlength{\parindent}{0pt}

These wave features, including mode localization, evanescent decay, quasi-normal-mode formation, and frequency-dependent scattering, naturally emerge from the way perturbations evolve in the curved spacetime of the wormhole. In this geometry, the redshift and shape functions define an effective potential that governs the propagation of scalar, electromagnetic, or gravitational waves near the throat. When the wave frequency lies below the height of this potential, the perturbations are exponentially damped, giving rise to evanescent or localized modes confined close to the throat. For higher frequencies, waves can partially transmit and reflect, producing resonant oscillations or quasi-normal modes characterized by complex frequencies that describe oscillation and damping rates. Physically, these quasi-normal modes represent the natural oscillatory response of the wormhole to external disturbances. Unlike black holes, whose event horizons irreversibly absorb outgoing radiation, traversable wormholes have no horizons. Waves can propagate between the two asymptotically flat regions, leading to interference effects across the throat. This repeated reflection can create echo-like or frequency-dependent modulations in the emitted signal, potentially detectable in precise gravitational-wave measurements. Several recent studies have shown that such echoes can serve as telltale signatures distinguishing horizonless compact objects from black holes \cite{CardosoTO-1, BuenoTO, KonoplyaTO, BronnikovTO, Blázquez-SalcedoTO, CardosoTO-2}. Moreover, the interference between multiple reflected waves near the throat can enhance mode localization and generate standing-wave patterns that depend sensitively on the redshift profile. The resulting frequency selectivity resembles an effective refractive behavior of the spacetime, where the curvature and matter content influence the propagation speed and intensity of specific modes \cite{AkhmedovTO, LinTO}. In this sense, the throat geometry, redshift gradient, and surrounding halo structure jointly determine how information about the internal wormhole configuration becomes encoded in observable wave phenomena.

\vspace{0.05cm}
\setlength{\parindent}{0pt}

This framework establishes a predictive methodology linking wormhole throat geometry, halo structure, and redshift effects to observable wave phenomena, including mode localization, evanescent decay, quasi-normal-mode formation, and frequency-dependent scattering. Its fully general nature allows the study of arbitrary halo profiles and redshift functions without relying on perturbative approximations, providing a robust foundation for analytical and numerical investigations of scalar-wave dynamics.

\vspace{0.05cm}
\setlength{\parindent}{0pt}

The correspondence between null geodesics and scalar-wave propagation emerges in the high-frequency (geometric-optics) limit, where the curvature-dependent terms become negligible in the effective refractive index $n_{\rm eff}(r,\omega)$. In this regime, the wave turning points coincide with the classical radial turning points of null rays, and extrema of $n_{\rm eff}$ reproduce the photon sphere condition. At finite frequencies, deviations from null geodesics arise due to the curvature-dependent contributions in $n_{\rm eff}$, giving rise to evanescent regions, quasi-bound states, and tunneling phenomena, which have no direct analog in classical ray optics. This establishes a direct and quantitative link between classical trajectories and wave-optical effects in wormhole spacetimes.

\vspace{0.05cm}
\setlength{\parindent}{0pt}

Within this correspondence, our explicit analysis of representative redshift profiles demonstrates that photon-sphere formation and shadow production are governed entirely by the radial structure of the redshift function $\Phi(r)$. By extremizing the function $r^{2}e^{-2\Phi(r)}$, which controls both the degeneracy of null-geodesic turning points and the trapping of high-frequency waves in the geometric-optics limit, we have shown that zero-redshift and cored redshift configurations admit no finite-radius extrema in the physical domain $r\ge r_{0}$. As a consequence, these geometries support neither circular null orbits nor shadows, independently of the choice of shape function and even when negative but regular redshift cores are allowed. In contrast, Teo-type redshift functions of the form $\Phi(r)=-a/r$ generate a unique unstable photon sphere located at $r_{\mathrm{ph}}=a$, with a critical impact parameter $b_{c}=a\,e$, thereby producing a well-defined and potentially observable shadow edge for asymptotic observers. This conclusion is further reinforced from the wave-optical perspective: in the geometric-optics limit, extrema of the effective refractive index coincide exactly with the null-geodesic trapping condition, while finite-frequency corrections may introduce evanescent regions but cannot generate photon spheres in the absence of a classical extremum of $r^{2}e^{-2\Phi(r)}$. These results establish unambiguously that shadow formation in wormhole spacetimes is dictated by the detailed radial behavior of the redshift function rather than by topology alone.

\vspace{0.05cm}
\setlength{\parindent}{0pt}

From an observational perspective, the refractive index analogy provides a rigorous optical framework to interpret scalar-wave dynamics in terms of tunneling through classically forbidden regions, mode confinement near the throat, and guided propagation shaped by halo-induced variations in $n_{\rm eff}(r,\omega)$. In principle, these optical analogues suggest potential observable imprints in both electromagnetic and gravitational wave signals. Low-frequency waves encountering evanescent regions may generate frequency-dependent echoes, while spatial variations in the halo density and redshift profile can induce phase shifts, wavefront distortions, and selective suppression of higher angular-momentum modes. Quasi-bound states localized near the throat can produce resonant enhancement of specific frequencies, leading to distinct spectral features. Quantitative modeling of $n_{\rm eff}(r,\omega)$ thus enables predictions for attenuation, scattering, and resonance patterns, offering a systematic approach to identify indirect observational signatures of wormholes embedded within realistic dark matter halos. 

\vspace{0.05cm}
\setlength{\parindent}{0pt}

In the context of gravitational wave observations, these phenomena imply that spacetime perturbations traversing a halo-embedded wormhole could exhibit delayed echoes, amplitude modulations, phase shifts, and spectral filtering analogous to the scalar-wave behavior. Precise measurements of GW waveforms, particularly in the low-frequency regime where quasi-bound states dominate, could reveal these signatures. High-resolution electromagnetic observations and laboratory analogues that simulate tailored refractive-index landscapes may provide complementary evidence, establishing a multi-messenger framework for detecting the imprints of exotic wormhole geometries in realistic astrophysical environments.

\vspace{0.05cm}
\setlength{\parindent}{0pt}

On the other hand, one may construct a transformation-optics analogue of this spacetime by isolating the spatial three-metric $\gamma_{ij}$, whose nonzero components are \cite{TO-1,TO-2,TO-3}
\begin{equation}
\gamma_{rr} = \frac{1}{1 - \epsilon(r)/r}, \quad 
\gamma_{\theta\theta} = r^2, \quad 
\gamma_{\varphi\varphi} = r^2 \sin^2\theta.
\end{equation}
The determinant of the spatial metric is therefore
\begin{equation}
\det \gamma = \frac{r^4 \sin^2\theta}{1 - \epsilon(r)/r},
\end{equation}
in contrast to flat space, where $\det \gamma_0 = r^4 \sin^2\theta$, yielding the volume element ratio
\begin{equation}
\sqrt{\frac{\det \gamma}{\det \gamma_0}} = \frac{1}{\sqrt{1 - \epsilon(r)/r}}.
\end{equation}
The inverse metric reads
\begin{equation}
\gamma^{rr} = 1 - \frac{\epsilon(r)}{r}, \quad 
\gamma^{\theta\theta} = \frac{1}{r^2}, \quad 
\gamma^{\varphi\varphi} = \frac{1}{r^2 \sin^2\theta},
\end{equation}
satisfying the canonical relation $\gamma^{ik} \gamma_{kj} = \delta^i_j$. According to the formalism of transformation optics, the effective permittivity and permeability tensors are \cite{TO-1,TO-2,TO-3}
\begin{equation}
\varepsilon^{ij} = \mu^{ij} = \sqrt{\frac{\det \gamma}{\det \gamma_0}} \, \gamma^{ij},
\end{equation}
which, upon lowering indices and including the temporal redshift factor $e^{\Phi(r)}$, yielding the covariant tensors
\begin{equation}
\begin{split}
 &\varepsilon_{rr} = \mu_{rr} = \frac{\sqrt{1 - \epsilon(r)/r}}{e^{\Phi(r)}}, \quad
\varepsilon_{\theta\theta} = \mu_{\theta\theta} = \frac{r^2}{\sqrt{1 - \epsilon(r)/r} \, e^{\Phi(r)}}, \\
&\varepsilon_{\varphi\varphi} = \mu_{\varphi\varphi} = \frac{r^2 \sin^2\theta}{\sqrt{1 - \epsilon(r)/r} \, e^{\Phi(r)}}. \label{cov-tensors}
\end{split}
\end{equation}
These tensors are inherently radially inhomogeneous and anisotropic. The local optical path lengths are governed by the redshift function $\Phi(r)$, while the curvature and confinement of quasi-bound states are controlled by the shape function $\epsilon(r)$. In particular, a Teo-type redshift combined with an NFW halo yields pronounced localization near the throat, whereas a cored halo with BEC oscillations produces multiple, spatially separated partial confinement regions. For zero-redshift scenarios with a PI halo, propagation is nearly uniform across the spatial domain, particularly for relatively high-frequency waves. Experimentally, these tensors in \eqref{cov-tensors} could, in principle, be implemented using metamaterials with radially graded anisotropic permittivity and permeability. Concentric layering strategies, achieving $\varepsilon_{rr}$ and $\varepsilon_{\theta\theta} = \varepsilon_{\varphi\varphi}$ profiles, may be implemented via three-dimensional gradient-index architectures, photonic crystals with radial lattice variations, or microwave waveguides featuring radially modulated thickness \cite{TO-1,TO-2,TO-3}. The quasi-bound states predicted by the theoretical framework could be probed through transmission and reflection spectroscopy, revealing curvature-induced resonances, while near-field scanning or local probe measurements can resolve spatial field distributions and confirm the predicted anisotropy. Temporal observables, such as phase delay or time-of-flight measurements, allow direct characterization of effective redshift effects induced by $\Phi(r)$, providing quantitative validation of the metric-metamaterial correspondence \cite{TO-4}. Nonetheless, several challenges remain. Discretizing the continuous profiles $\epsilon(r)$ and $\Phi(r)$ into finite concentric layers introduces deviations from the ideal metric, especially near regions of strong curvature or steep redshift gradients. Material losses, imperfect anisotropy, and bandwidth limitations may attenuate or smear the quasi-bound state signatures. Furthermore, achieving high-fidelity three-dimensional index gradients at optical frequencies is experimentally demanding, with potential parasitic scattering or mode-mixing effects \cite{TO-1,TO-2,TO-3,TO-4}. Despite these limitations, the proposed framework offers a systematically tunable, fully analytic platform for exploring wormhole-inspired and dark-matter halo-mimicking optical phenomena, allowing controlled separation of throat geometry, curvature effects, and redshift contributions.

\vspace{0.02cm}
\setlength{\parindent}{0pt}

Taken together, these results show that shadowless wormholes arise naturally in physically well-motivated cored redshift configurations, including those with negative but regular cores, whereas shadow-casting wormholes require sufficiently steep redshift gradients capable of supporting unstable null trapping. This distinction provides a clear and physically grounded observational discriminator between different classes of horizonless compact objects.

\end{document}